\input{aipcheck}

\documentclass[
    ,final            
  ]
  {aipproc}

\layoutstyle{6x9}

\begin{document}

\title{Three-Field Potential for Soft-Wall AdS/QCD}


\author{Sean P. Bartz}{
  address={116 Church St SE, Minneapolis, MN 55455}
}

\author{Joseph I. Kapusta}{
  address={116 Church St SE, Minneapolis, MN 55455}
}

\begin{abstract}
The Anti-de Sitter Space/Conformal Field Theory (AdS/CFT) correspondence may offer new and useful insights into the non-perturbative regime of strongly coupled gauge theories such as Quantum Chromodynamics (QCD). Soft-wall AdS/QCD models have reproduced the linear Regge trajectories of meson spectra by including background dilaton and chiral condensate fields. Efforts to derive these background fields from a scalar potential have so far been unsuccessful in satisfying the UV boundary conditions set by the AdS/CFT dictionary while reproducing the IR behavior needed to obtain the correct meson spectra. We present work toward a three-field scalar potential that includes the dilaton field and the chiral and glueball condensates. This model is consistent with background fields that yield linear trajectories for the meson spectra and the correct mass-splitting between the vector and axial-vector mesons.
\end{abstract}

\maketitle


\section{Introduction and Motivation}

Quantum chromodynamics has been well tested for high-energy collisions, where perturbation theory is applicable. However, at hadronic scales, the interaction is non-perturbative, requiring a new theoretical model. The Anti-de Sitter Space/Conformal Field Theory (AdS/CFT) correspondence establishes a connection between $n$-dimensional Super-Yang Mills Theory and a weakly-coupled gravitational theory in $n+1$ dimensions \cite {maldacena}. Phenomenological models inspired by this correspondence are known as AdS/QCD, and have succeeded in capturing some features of QCD \cite{stephanov-katz-son}. 

 Quark confinement in QCD sets a scale that is encoded in a cut-off of the fifth dimension in the AdS theory. Soft-wall models use a dilaton as an effective cut-off to keep the meson fields from penetrating too deeply into the bulk. The simplest soft-wall models use a quadratic dilaton to recover the linear Regge trajectories, \cite{stephanov-katz-son} while models that modify the UV behavior of the dilaton more accurately model the ground state masses. In this work, we use the meson action from \cite{gherghetta-kelley}. 
 \begin{equation}
 S_{5}=\int d^{5}x\sqrt{-g}e^{-\phi(z)}Tr\left[|DX|^{2}+m_{X}^{2}|X|^{2}+\kappa|X|^4+\frac{1}{4g_{5}^{2}}(F_{L}^{2}+F_{R}^{2})\right]. \label{eq:mesonaction}
\end{equation}
Where $\phi$ is the dilaton, $X$ is the scalar field, and $F_{L,R}$ contain the vector and axial vector gauge fields. Please see \cite{gherghetta-kelley, bartz-pions} for details on the field content, the equations of motion, and the numerical calculation of the meson spectra.


We begin examining the background equations for the dilaton and chiral condensate fields ($\phi$ and $\chi$), using the action proposed in \cite{Springer2010}. The chiral condensate is identified as the vacuum expectation value of the scalar field in (\ref{eq:mesonaction}). In the Einstein frame, the action for the background fields reads
\begin{equation}
S_E=\frac{1}{16\pi G_5}\int d^{5}x\sqrt{-g_E}\left(R_E-\frac{1}{2}\partial_\mu\phi\partial^\mu\phi-\frac{1}{2}\partial_\mu\chi\partial^\mu\chi-V(\phi,\chi)\right)
\end{equation}
One background equation does not depend on the potential,
\begin{equation}
6a\phi''(z)+[\phi'(z)]^{2}(6a^{2}-1)-[\chi'(z)]^{2}+\frac{12a\phi'(z)}{z}=0.  \label{V-indep1}
\end{equation}
Examining (\ref{V-indep1}), and requiring the IR behavior that $\phi=\lambda z^2$ and $\chi = \Gamma  z$, determines $a=1/\sqrt{6}$ \cite{Springer2010}. This determines a relationship between $\lambda$, the parameter that sets the slope of the Regge trajectories, and $\Gamma$, which is related to the mass-splitting between the vector and axial vector mesons \cite{gherghetta-kelley}
\begin{equation}
\Delta m^2 = \lim_{z\rightarrow\infty} \frac{g_5^2\chi^2}{z^2}=g_5^2\Gamma^2=g_5^2 6^{3/2}\lambda
\end{equation}
 Using the experimental value of $\lambda$  gives a value of $\Delta m^2$ that is $\sim50$ times too large.

\section{Three-Field Model}

One proposed solution to this problem is to add another scalar field to the action for the background fields, $G$,  dual to the glueball. By this association, the UV boundary condition is $\lim_{z\rightarrow 0} G(z)=g_o z^4$, where $g_o$ is the gluon condensate. To maintain linear confinement, $G\sim z$ in the IR. The background equations that result are
\begin{eqnarray}
\chi'^{2}+G'^{2}&=&\frac{\sqrt{6}}{z^{2}}\frac{d}{dz}\left(z^{2}\phi'\right) \label{background1} \\
V+12&=&\frac{\sqrt{6}}{2}z^{2}\phi''-\frac{3}{2}(z\phi')^{2}-3\sqrt{6}\phi' \\
\frac{\partial V}{\partial\phi}&=&3z\phi' \label{background3} \\
\frac{\partial V}{\partial\chi}&=&z^{2}\chi''-3z\chi'\left(1+\frac{z\phi'}{\sqrt{6}}\right) \label{background4} \\
\frac{\partial V}{\partial G}&=&z^{2}G''-3zG'\left(1+\frac{z\phi'}{\sqrt{6}}\right). \label{background5}
\end{eqnarray}
Because the potential depends on $z$ only through the fields, we can eliminate one of (\ref{background3}), (\ref{background4}), or (\ref{background5}).

Examining (\ref{background1}) in the IR limit, we see that we can adjust the coefficients of the power-law behavior of the dilaton and chiral condensate fields independently, matching them to their experimental values.


We seek a parametrization for the chiral and glueball fields that is consistent with an expression for the dilaton that can be written without any special functions. In addition, the parametrization must yield meson spectra that match well with experiment. The simplest expressions that match these criteria were found to be 
\begin{equation}
G'(z)=\frac{A}{B^3}\left(1-e^{-Bz}\right)^3, \quad \quad
\chi'(z)=\frac{\alpha}{\beta^2}\left(1-e^{-\beta z}\right)^2.
\end{equation}
 The above parameters are defined: $\alpha=\sigma/3, \, \alpha/\beta^2=\Gamma, \, A=g_o/4. $  $B$ is set by ensuring (\ref{background1}) is satisfied in the IR limit.

\begin{figure}
  \includegraphics[height=.45\textheight]{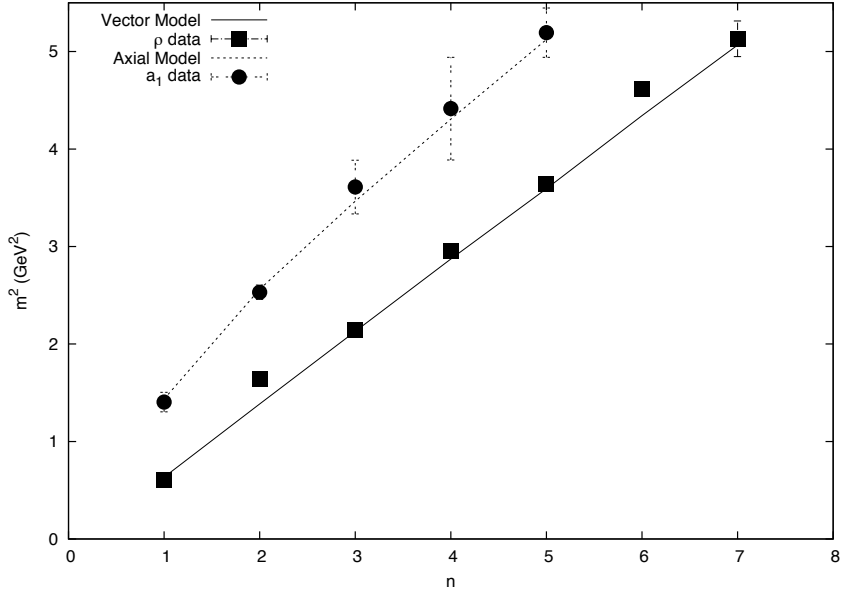} \label{fig:only}
  \caption{The experimental $\rho$ and $a_1$ meson masses \cite{PDG} are fit well by the eigenvalues found using this parametrization.}
\end{figure}

The values of the parameters, as determined by a least-squares fitting to the $\rho$ and $a_1$ spectra, are: $\sigma = (0.375\, \rm{GeV})^3,\,  g_o=(1.5\,\rm{GeV})^4,\,\Gamma = 0.25\, \rm{GeV}, \, \lambda = (0.428\,\rm{GeV})^2$. The meson spectra that result from this parametrization are shown in Figure \ref{fig:only}.

%




\bibliographystyle{aipproc}   

\bibliography{seanbartz}

\IfFileExists{\jobname.bbl}{}
 {\typeout{}
  \typeout{******************************************}
  \typeout{** Please run "bibtex \jobname" to optain}
  \typeout{** the bibliography and then re-run LaTeX}
  \typeout{** twice to fix the references!}
  \typeout{******************************************}
  \typeout{}
 }

\end{document}